\documentclass[
  prmaterials,
  reprint,
  groupedaddress,
  amsmath,
  floatfix,
]{revtex4-2}

\usepackage[T1]{fontenc}
\usepackage{libertinus}
\usepackage{microtype}

\usepackage{bm}
\usepackage{siunitx}

\usepackage{graphicx}
\graphicspath{{figures/}}
\usepackage{subfig}

\usepackage{booktabs}
\usepackage{multirow}

\usepackage{xr-hyper}
\usepackage{url}
\usepackage{hyperref}
\usepackage{cleveref}
\usepackage{xcolor}
\usepackage{soul}

\begin{document}

\title{Benchmarking average atom potentials derived from atomic cluster expansions}

\author{Deepak Somani}
\affiliation{Laboratory of Materials Design and Simulation (MADES), Institute of Materials, \'{E}cole Polytechnique F\'{e}d\'{e}rale de Lausanne}

\author{Lorenzo Piersante}
\affiliation{Laboratory of Materials Design and Simulation (MADES), Institute of Materials, \'{E}cole Polytechnique F\'{e}d\'{e}rale de Lausanne}

\author{Anirudh Raju Natarajan}
\email{anirudh.natarajan@epfl.ch}
\affiliation{Laboratory of Materials Design and Simulation (MADES), Institute of Materials, \'{E}cole Polytechnique F\'{e}d\'{e}rale de Lausanne}

\date{\today}

\begin{abstract}
  Average atom (A-atom) potentials provide a mean-field description of a chemically disordered alloy and are used to predict the properties of solid solutions without short-range order. Such potentials are usually averaged from an existing interatomic potential and are therefore only as accurate as the parent model. Accurate interatomic potentials are themselves difficult to parameterize and can require large training datasets. Here we benchmark a recently developed formalism that computes an exact A-atom potential directly from a linear atomic cluster expansion (ACE). We first fit a linear ACE to Fe-W data generated with an embedded atom method (EAM) potential. The resulting A-atom potential reproduces the properties of the disordered phase computed from an explicit random supercell and from a conventional A-atom potential averaged from the same EAM potential. We then fit a linear ACE to energies and forces computed from electronic structure calculations for about 1500 small Mo-Nb structures with an average of 7 atoms per structure. The A-atom potential derived from this ACE reproduces the DFT elastic constants, lattice parameter, mixing enthalpy, and Bain path of special quasirandom structures. Of the three chemical site bases, only the occupation basis also reproduces the DFT surface energies of the alloy. For Mo-Nb, the A-atom potential also predicts that the ideal solution entropy outweighs the destabilizing vibrational contribution to the finite-temperature free energy of the disordered phase. These benchmarks show that the properties of disordered alloys can be recovered from small training datasets when the chemical site basis is chosen carefully.
\end{abstract}

\maketitle

\section{Introduction}
Chemically disordered multicomponent solid solutions are of current interest for a range of engineering applications. Their thermodynamic, kinetic, and mechanical response is determined by the distribution of chemical elements across the underlying parent crystal structure. Modeling chemically disordered alloys is challenging, as any single configuration samples only a small part of the phase space an alloy explores. For instance, thermodynamic properties can be approximated with special quasirandom structures (SQS)\cite{zunger_1990_specialquasirandom}. These are supercells whose short-range correlations are constructed to match those of a random alloy. SQS have also been used to compute inputs to models of dislocation properties and of the susceptibility of an alloy to cracking\cite{liyanage_2026_crackingcase}. A complementary route coarse-grains over the chemical degrees of freedom and replaces every species with a single effective atom. Average atom (A-atom) potentials\cite{varvenne_2016_averageatominteratomic,smith_1989_applicationmolecular} have been derived and validated for semi-empirical interatomic potentials based on the embedded atom method (EAM). They have since been used to model dislocation behavior in concentrated alloys\cite{varvenne_2017_solutestrengthening,maresca_2020_theoryscrew,maresca_2026_strengthscrew,ghafarollahi_2022_screwcontrolledstrength} and to build predictive models of strength. EAM potentials are themselves approximate and can show unphysical behavior\cite{daw1993,moller2018,zimmerman2000}. Consequently, an A-atom potential built from such a model inherits those errors.

Interatomic potentials fitted directly to electronic structure calculations offer a more reliable atomistic model, but the chemical average that produces an A-atom potential is not as readily available as it is for the embedded atom method. A recent study\cite{piersante_2026_ACEchemical} derived an exact methodology for computing the A-atom potential from a linear atomic cluster expansion (ACE)\cite{drautz_2019_atomiccluster,drautz_2020_atomiccluster}. The derivation treats the chemical degrees of freedom explicitly and gives an analytical A-atom potential for a perfectly random alloy. The resulting A-atom potential retains the computational benefits of the parent ACE, treating the disordered alloy as an effective single-element material and preserving linear scaling. Any linear ACE can therefore be converted into an A-atom potential at a specific composition without refitting. The mapping has so far been demonstrated on mixing enthalpies, where a modest training set of electronic structure calculations was sufficient to reproduce those of the disordered phase\cite{piersante_2026_ACEchemical}. Mixing enthalpies, however, probe only one part of the energy landscape. Whether an A-atom potential obtained in this way also reproduces point, line, and planar defect energies, which control the kinetic and mechanical response of the alloy, has not been established.

In this study we provide these benchmarks. After a short overview of the formalism, we test it on two prototypical alloy systems. The first test uses an Fe-W EAM potential as a reference. We fit a linear ACE to data generated with that potential and compare the resulting A-atom potential against explicit random decorations of the same alloy. We also compare it against the A-atom potential obtained directly from the EAM parameters following the approximation of Varvenne \textit{et al.}\cite{varvenne_2016_averageatominteratomic}. The two A-atom potentials agree across the thermodynamic, kinetic, and mechanical properties we examine. The second test applies this construction to a binary Mo-Nb alloy, where the linear ACE is parameterized on electronic structure calculations rather than on a semi-empirical potential. We compare three chemical site bases for the linear ACE. With the occupation basis, the resulting A-atom potential reproduces the alloy surface energies of Liyanage \textit{et al.}\cite{liyanage_2026_crackingcase} and the composition dependence of the unstable stacking fault energies. The Chebyshev and ACE bases deviate from these predictions for the surface and vacancy energies at intermediate compositions. The training dataset contains no defect configurations at these compositions, so the A-atom potential must extrapolate there. With the larger Fe-W training dataset, the three bases give indistinguishable predictions. Taken together, these benchmarks show that a physically representative description of a disordered alloy can be parameterized from a small set of calculations, drawn either from electronic structure or from an existing interatomic potential such as a universal foundation model. These A-atom potentials also provide a reference against which a real alloy, with its short-range order and local relaxations, can be compared to quantify the consequences of departures from ideal randomness.

\section{Methods}
\label{sec:methods}

\subsection{Deriving A-atom potentials from linear ACE models}
\label{sec:a-atom-derivation}

The contribution of each site to the total energy of a crystal can be expanded in a basis of local chemical and geometric descriptors\cite{drautz_2019_atomiccluster,drautz_2020_atomiccluster}:
\begin{equation}
  \label{eq:ACE_site_energy}
  E_{i} = \sum_{\boldsymbol{\eta} \boldsymbol{\nu}} c_{\boldsymbol{\eta} \boldsymbol{\nu}} B_{\boldsymbol{\eta} \boldsymbol{\nu},i}
\end{equation}
where $\boldsymbol{\eta}$ is a tuple that collects the chemical indices of the basis functions entering a descriptor, $\boldsymbol{\nu}$ similarly tracks the radial and angular basis labels, $B_{\boldsymbol{\eta} \boldsymbol{\nu},i}$ describes the structural and chemical environments around site $i$, and $c_{\boldsymbol{\eta} \boldsymbol{\nu}}$ are learnable expansion coefficients. Each descriptor $B_{\boldsymbol{\eta} \boldsymbol{\nu},i}$ is a symmetrized product of the atomic densities $A_{\eta,nlm}$:
\begin{equation}
  B_{\boldsymbol{\eta} \boldsymbol{\nu},i} = \sigma_{\eta_{0}}(\mu_{i}) \sum_{\boldsymbol{m}} C_{\boldsymbol{l} \boldsymbol{m} \boldsymbol{L} \boldsymbol{m}}^0 \prod_{(\eta,nlm) \in (\boldsymbol{\eta},  \boldsymbol{\nu})} A_{\eta,nlm}
\end{equation}
in which the generalized Clebsch--Gordan coefficients $C_{\boldsymbol{l} \boldsymbol{m} \boldsymbol{L} \boldsymbol{m}}^0$ enforce the symmetry invariances of $B_{\boldsymbol{\eta} \boldsymbol{\nu},i}$ and $\sigma_{\eta_{0}}$ is the chemical site basis function for the species $\mu_{i}$ that occupies central site $i$. The atomic densities form the $A$-basis and collect the contribution of every neighbor $j$ of site $i$:
\begin{equation}
  A_{\eta,nlm} = \sum_{j}\sigma_{\eta}(\mu_{j}) R_{nl}(r_{ji}) Y_l^m(\hat{r}_{ji})
\end{equation}
where $\eta$ indexes the chemical basis, $R_{nl}$ are radial basis functions, $Y_l^m$ are spherical harmonics, and $r_{ji}$ and $\hat{r}_{ji}$ are the distance and the unit vector between sites $j$ and $i$.

The chemical, radial, and angular basis functions can each be chosen in several ways. Spherical harmonics are the most convenient angular basis, since they make the symmetry invariances of the $B$-basis functions straightforward to enforce\cite{drautz_2019_atomiccluster,drautz_2020_atomiccluster}. The radial basis can be drawn from any complete set, such as Chebyshev polynomials or radial Bessel functions\cite{drautz_2019_atomiccluster, bigi_2026_metatensormetatomic}. In this study we use real spherical harmonics as defined in Ref.~\cite{bigi_2023_fastevaluation} and radial Bessel functions, and do not explore the effect of other radial bases. Instead, we focus on three chemical site bases that can be used to set up a linear ACE. Chebyshev polynomials and indicator functions both form complete chemical site bases in the on-lattice cluster expansion\cite{sanchez_1984_generalizedcluster}, and the indicator functions are traditionally referred to as the occupation basis. Conventional ACE potentials instead adopt a unit-vector (one-hot) site basis, in which the occupation basis is augmented with one fictitious chemical species, such as a vacancy, that is never included in the parameterization or evaluation of the model. For a binary alloy, the three basis sets, labeled $\boldsymbol{\sigma}_{\textrm{Cheby}}$, $\boldsymbol{\sigma}_{\textrm{Occ}}$, and $\boldsymbol{\sigma}_{\textrm{ACE}}$, have site basis functions given by:
\begin{equation}
  \label{eq:chemical_site_basis}
  \begin{gathered}
    \boldsymbol{\sigma}_{\textrm{Cheby}} = \begin{pmatrix} 1 \\ 2\delta_{\mu_{i},B}-1 \end{pmatrix}, \quad
    \boldsymbol{\sigma}_{\textrm{Occ}} = \begin{pmatrix} 1 \\ \delta_{\mu_{i},B} \end{pmatrix}, \\
    \boldsymbol{\sigma}_{\textrm{ACE}} = \begin{pmatrix} 1 \\ \delta_{\mu_{i},A} \\ \delta_{\mu_{i},B} \end{pmatrix}
  \end{gathered}
\end{equation}
where $\delta_{\mu_{i},A}$ evaluates to 1 if species $A$ occupies site $i$ and 0 otherwise. The Chebyshev and occupation bases are chemically complete for a binary alloy, while the conventional ACE basis is overcomplete. A constant site basis function enables the construction of a hierarchical cluster expansion. Overcompleteness poses no formal difficulty for a linear ACE model.

A linear ACE is parameterized on a training database of energies and forces computed for a structurally and chemically diverse set of configurations. The site-centric descriptors can be evaluated with existing codes such as \texttt{pacemaker}\cite{bochkarev_2022_efficientparametrization}, \texttt{torch-spex}\cite{bigi_2026_metatensormetatomic}, and \texttt{ACEsuit}\cite{witt_2023_acepotentialsjljuliaa}. The coefficients $c_{\boldsymbol{\eta} \boldsymbol{\nu}}$ are then obtained by linear regression, using ordinary least squares in the data-rich regime and regularized least squares with a Lasso or Ridge penalty otherwise. Typical ACE potentials for a pure element or a binary alloy require 10,000 to 20,000 structures in the training dataset to reproduce the properties of interest with sufficient accuracy.

An A-atom potential is obtained exactly from a linear ACE by evaluating the $B$-basis functions in an equivalent mean-field alloy, in which every site is occupied by an averaged element of the same composition as the alloy. The A-atom site energy $\overline{E}_{i}$ is then:
\begin{equation}
  \label{eq:ACE_A_atom}
  \overline{E}_{i} = \sum_{\boldsymbol{\eta} \boldsymbol{\nu}} {c}_{\boldsymbol{\eta} \boldsymbol{\nu}} \langle B_{\boldsymbol{\eta} \boldsymbol{\nu},i}\rangle
\end{equation}
where the averaged descriptors $\langle B_{\boldsymbol{\eta} \boldsymbol{\nu},i}\rangle$ are computed analytically from the alloy composition and structure. The averaged descriptors separate into a leading conventional ACE term and several self-interaction terms\cite{piersante_2026_ACEchemical}:
\begin{equation}
  \label{eq:averaged_B_basis}
  \langle B_{\boldsymbol{\eta} \boldsymbol{\nu}} \rangle = \langle \sigma_{\eta_0} \rangle \left(\prod_{k = 1}^K \langle \sigma_{\eta_k} \rangle \right) B_{\boldsymbol{\nu}} + \sum_{\boldsymbol{b} \in \mathcal{P}(\boldsymbol{\eta})} S_{\boldsymbol{b}}^{\text{chem}} \sum_{\tilde{\boldsymbol{\nu}}(\boldsymbol{b})} C_{\tilde{\boldsymbol{\nu}}}^{\text{ang}} B_{\tilde{\boldsymbol{\nu}}}^{\text{self}}
\end{equation}
where $\langle \sigma_{\eta} \rangle = \sum_{\mu} x_{\mu} \sigma_{\eta}(\mu)$ is the composition-weighted average of a chemical site basis function over the species $\mu$ present at concentrations $x_{\mu}$, $\eta_{0}$ is the chemical index of the central site and $\eta_{k}$ that of the $k$-th bond, and $K$ is the number of bonds in the descriptor $B_{\boldsymbol{\nu}}$. The set $\mathcal{P}(\boldsymbol{\eta})$ contains the set partitions of the $K$ bonds, excluding the case of totally distinct bonds, and each partition $\boldsymbol{b}$ specifies which bonds coincide on the same neighbor. For each partition, $S_{\boldsymbol{b}}^{\text{chem}}$ is the chemical factor and $C_{\tilde{\boldsymbol{\nu}}}^{\text{ang}}$ is the $l$-dependent angular prefactor attached to each permitted self-interacting $B$-basis function $B_{\tilde{\boldsymbol{\nu}}}^{\text{self}}$. A tilde marks the multi-indices of self-interacting clusters, and the site index $i$ is dropped from \cref{eq:averaged_B_basis} for readability. The $B_{\boldsymbol{\nu}}$ are rotationally invariant descriptors built from the standard ACE $A$-basis and evaluated as though every site were occupied by the average atom. The $B_{\tilde{\boldsymbol{\nu}}}^{\text{self}}$ are likewise rotationally invariant and evaluated in the same effective structure, but they combine the standard $A$-basis with an augmented $A$-basis in which radial functions are repeated on the same bond. Analytical expressions for the chemical and $l$-dependent factors of pair, three-body, and four-body interactions are derived in Ref.~\cite{piersante_2026_ACEchemical}. We restrict ourselves to linear ACE potentials containing up to three-body interactions, which we find sufficient for the alloys considered here.

An advantage of deriving an A-atom potential from a linear ACE is that the chemical and $l$-dependent factors in \cref{eq:averaged_B_basis} are computed analytically and \textit{a priori}. Beyond the standard descriptor $B_{\boldsymbol{\nu}}$, the only quantity that must be evaluated during a simulation is the self-interaction $B_{\tilde{\boldsymbol{\nu}}}^{\text{self}}$, which can be computed while preserving linear scaling with the number of neighbors. \Cref{app:average_zbl} applies the same composition averaging to the Ziegler--Biersack--Littmark repulsion used at short interatomic distances.

\subsection{Training dataset}
\label{sec:training-dataset}

Separate training datasets were generated for the Fe-W and Mo-Nb binary systems, each containing the energies and forces of a chemically and structurally diverse set of configurations. Energies and forces were evaluated with the EAM potential of Zhou \textit{et al.}\cite{zhou_2004_misfitenergyincreasingdislocations} for Fe-W and with electronic structure calculations for Mo-Nb. In both systems, symmetrically distinct chemical decorations were enumerated over several parent crystal structures with the algorithm of Hart and Forcade\cite{hart_2008_algorithmgenerating,hart_2009_generatingderivative}.

For Fe-W, all symmetrically distinct decorations containing up to six atoms were enumerated on the bcc, hcp, and $\omega$ structures and on ABAC-type stackings of triangular layers. Each bcc decoration was strained along the symmetry-breaking directions defined by the strain order parameters of a finite-strain Hencky metric\cite{thomas_2017_explorationnonlinear,barsch_1984_twinboundaries}. A small number of larger cells were added for pure Fe and W, comprising a single vacancy in a $3\times 3\times 3$ supercell of the conventional bcc cell, and bcc slabs with $(111)$, $(110)$, and $(100)$ surfaces. All structures except the strained ones were relaxed with the EAM potential, and the energies and forces of every configuration along each relaxation trajectory were added to the training dataset. The Fe-W training dataset comprised 7843 energies and the forces on all 42428 atoms in these configurations.

The Mo-Nb training dataset was constructed analogously, with fewer data points. Energies and forces were computed with \emph{Vienna Ab-initio Simulation Package} (VASP, ver. 6.3.0)\cite{kresse_ab_1993, kresse_ab_1994, kresse_efficiency_1996, kresse_ultrasoft_1999} using a plane-wave cutoff of 550\,\unit{\electronvolt} and PAW-PBE pseudopotentials\cite{perdew_generalized_1996, perdew_generalized_1997} treating 14 and 13 valence electrons for Mo and Nb, respectively, on k-point grids with a spacing of approximately \SI{0.11}{\per\angstrom}.

Symmetrically distinct Mo-Nb decorations were enumerated on the bcc, fcc, hcp, ABAC, and $\omega$ prototypes, giving 142 decorations in total (57 bcc, 57 fcc, 13 hcp, 9 ABAC, and 6 $\omega$). All 142 decorations were relaxed with respect to all structural degrees of freedom, and the energies and forces along each relaxation trajectory were added to the training dataset. Each bcc, fcc, and hcp decoration was then strained by $\pm 0.1$ along all six strain order parameter directions. Decorations that remained bcc after relaxation, as classified with the structure-matching algorithm of Thomas \textit{et al.}\cite{thomas_2021_comparingcrystal}, were distorted by a further $\pm 0.05$ along the same directions. These strain distortions formed a reference dataset of 2712 structures. Rather than including all of them, we selected 1273 structures from the reference dataset by iterative sampling based on the Mahalanobis distance in the ACE descriptor space, following Ref.~\cite{piersante_2026_machinelearning}. The energies and forces of the selected structures were computed with static electronic structure calculations.

Defect structures were generated for pure Mo and Nb. A single vacancy was introduced into a $4\times 4\times 4$ supercell of the conventional bcc cell, and the exchange of the vacancy with a nearest neighbor was sampled at three points along the migration path, up to the activated state. Slabs with $(111)$, $(110)$, and $(100)$ surfaces were also generated.
Static images along the screw dislocation glide path were computed in a 231-atom cell containing a screw dipole constructed with \textsc{Atomman}, with intermediate images taken from a preliminary ACE parameterization. An unstable stacking fault was created by applying a rigid $\langle 111 \rangle$ half-crystal shift on a $\{110\}$ plane. The final Mo-Nb training dataset comprised 1722 energies and the corresponding forces, drawn from 1464 calculations and their intermediate relaxation steps.

\subsection{Fitting the ACE potentials}
\label{sec:ace-fitting}

For each alloy system, linear ACE models were fitted with each of the three chemical site bases in \cref{eq:chemical_site_basis}. The descriptors and the A-atom potentials were constructed with \texttt{featomic} and \texttt{metatensor}\cite{bigi_2026_metatensormetatomic}, as described in Ref.~\cite{piersante_2026_ACEchemical}. All models include up to three-body interactions, and the Fe-W and Mo-Nb models differ only in their interaction cutoff radii and basis sizes. The Fe-W models used a cutoff radius of 6.0\,\unit{\angstrom}, with 8 and 6 radial functions for the pair and triplet terms, respectively, and spherical harmonics up to $l = 6$ for the triplet terms. The Mo-Nb models used a cutoff radius of 7.0\,\unit{\angstrom}, with 12 and 8 radial functions for the pair and triplet terms, respectively, and spherical harmonics up to $l = 8$. Coefficients were obtained by Ridge regression, and the regularization strength was selected by five-fold cross-validation.

\subsection{Free energy calculations}
\label{sec:free_energy_calculations}

An A-atom potential gives direct access to the phonon modes of the random alloy without short-range order. These phonons provide the vibrational contribution to the finite-temperature free energy, and hence to the phase stability of the disordered state. Since the A-atom potential describes a perfectly random alloy, the configurational entropy is exactly the ideal solution entropy.

We calculate the mixing free energy of Mo-Nb with the A-atom potential derived from the occupation basis, using the quasi-harmonic approximation as implemented in \texttt{Phonopy}\cite{togo2023,togo2023a}. In a quantum mechanical treatment of the ions, the vibrational free energy depends on the atomic masses. Different choices for the mass of the average atom, which depend on the intended application, then yield slightly different free energies\cite{ramirez2010}. This ambiguity disappears in the classical limit, where relative phase stability is mass independent. We therefore calculate quasi-harmonic free energies within classical statistical mechanics. We set the mass of the average atom to $m_{\text{Nb}}^{1 - x} m_{\text{Mo}}^{x}$, where $x$ is the molybdenum concentration. With this choice, the ideal-gas free energy of the A-atom crystal matches that of the random alloy at the same concentration.

\section{Benchmarking the A-atom potential}
\label{sec:benchmarking}

\subsection{Fe-W from a semi-empirical potential}
\label{sec:fe-w-benchmark}

\begin{figure*}
  \centering
  \subfloat[]{\label{subfig:few-c11}\includegraphics[width=0.28\textwidth]{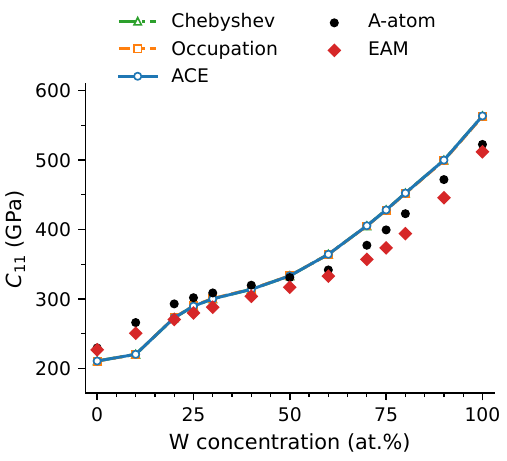}}\hfill
  \subfloat[]{\label{subfig:few-c12}\includegraphics[width=0.28\textwidth]{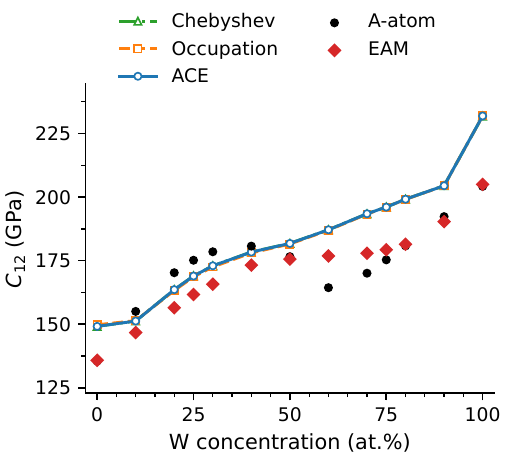}}\hfill
  \subfloat[]{\label{subfig:few-c44}\includegraphics[width=0.28\textwidth]{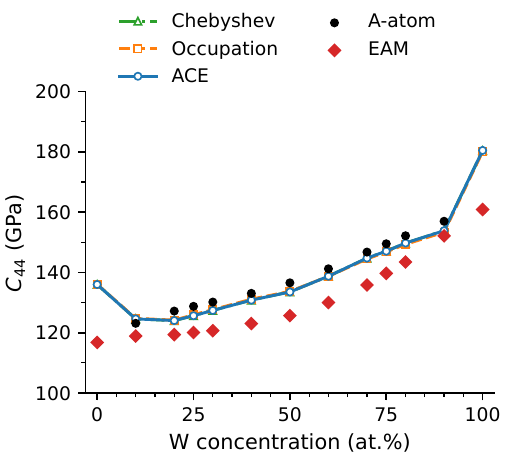}}\\
  \subfloat[]{\label{subfig:few-hmix}\includegraphics[width=0.28\textwidth]{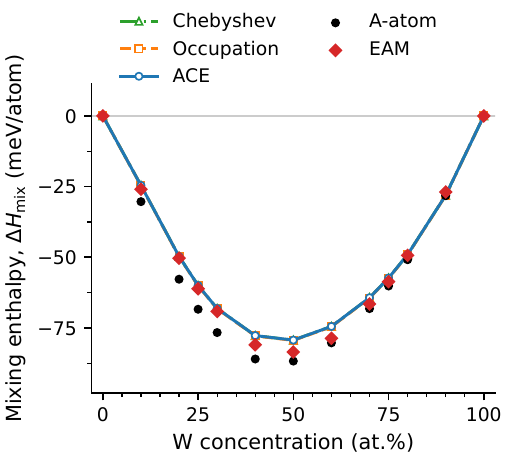}}\hfill
  \subfloat[]{\label{subfig:few-peierls}\includegraphics[width=0.28\textwidth]{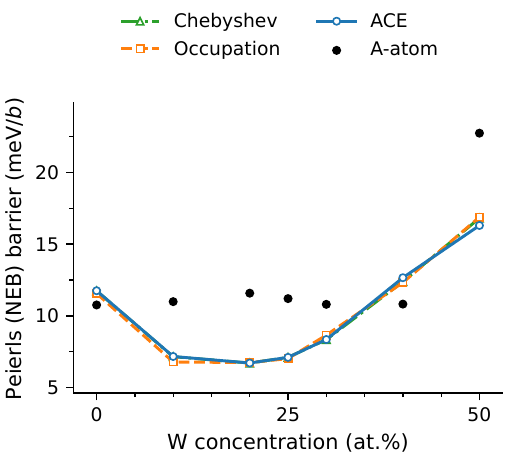}}\hfill
  \subfloat[]{\label{subfig:few-gamma110}\includegraphics[width=0.28\textwidth]{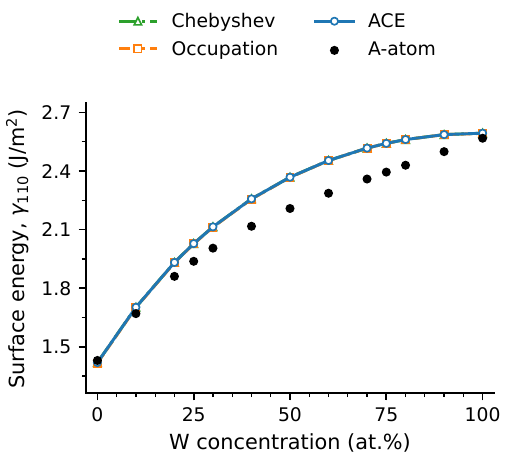}}
  \caption{Properties of the disordered Fe-W solid solution as a function of composition, computed with the A-atom potential derived from a linear ACE and with the A-atom potential obtained directly from the EAM parameters following the approximation of Varvenne \textit{et al.}\cite{varvenne_2016_averageatominteratomic}. Predictions of the linear ACE are shown for the Chebyshev, occupation, and ACE chemical site bases of \cref{eq:chemical_site_basis}. Black circles show the EAM A-atom potential, and red diamonds in (a)--(d) show values computed directly with the EAM potential in a large random supercell. (a) Elastic constant $C_{11}$. (b) Elastic constant $C_{12}$. (c) Elastic constant $C_{44}$. (d) Mixing enthalpy. (e) Peierls barrier for a $1/2\langle111\rangle$ screw dislocation with the dislocation line and Burgers vector oriented along the $\langle111\rangle$ direction. (f) Surface energy of the $(110)$ surface.}
  \label{fig:fe-w-benchmark}
\end{figure*}

We first compare the A-atom potential derived from a linear ACE against the A-atom potential obtained directly from the EAM potential with the approximation of Varvenne \textit{et al.}\cite{varvenne_2016_averageatominteratomic}. We refer to these as the ACE-derived and EAM A-atom potentials. Both descriptions of the disordered Fe-W solid solution are derived from the same EAM potential, since the linear ACE was parameterized on energies and forces generated with it. The comparison is therefore made against an established A-atom model of the disordered phase. Such a reference is not available once the training data come from electronic structure calculations. While the approximation of Varvenne \textit{et al.}\cite{varvenne_2016_averageatominteratomic} is well tested, it remains an approximation rather than an exact chemical average. Differences between the two A-atom potentials are expected, and do not by themselves indicate an error in either model. For the elastic constants and the mixing enthalpy, we also evaluate the EAM potential directly in a large random supercell of Fe and W across 3 different random decorations in a 6750 atom cell. The random supercell samples the disordered phase without any chemical averaging. Fitting to the Fe-W training dataset of \cref{sec:training-dataset} gave a linear ACE with root-mean-square errors of $6$\,meV/atom in energy and $47$\,meV/\si{\angstrom} in force. Energy and force errors alone do not establish that the derived A-atom potential describes the disordered alloy, so we compare material properties directly.

\Cref{fig:fe-w-benchmark} compares the predictions of the two A-atom potentials across six properties of the disordered solid solution. The Chebyshev, occupation, and ACE chemical site bases are indistinguishable on the scale of the figure for every property, so in this data-rich regime the choice of chemical site basis does not affect the predictions. The elastic constants are shown in \crefrange{subfig:few-c11}{subfig:few-c44}, and the two A-atom potentials follow the same trends across the full composition range. The largest difference in trend appears in $C_{12}$, where the ACE-derived A-atom potential varies smoothly with composition while the EAM A-atom potential falls from 180\,\unit{\giga\pascal} at 40\,at.\% W to 164\,\unit{\giga\pascal} at 60\,at.\% W before rising again. The random supercell shows no such decrease, and its $C_{12}$ increases monotonically with W concentration (\cref{subfig:few-c12}). At the pure elements, where the A-atom potential reduces to the parent EAM potential and the reference values are exact, the ACE-derived $C_{12}$ and $C_{44}$ are higher than the EAM values by $\approx 20$\,\unit{\giga\pascal}, which corresponds to a relative error of $\approx 10\%$. These errors arise from the fit of the linear ACE rather than from the chemical averaging, which is trivial at the pure elements. They could be reduced by reweighting the training data or enlarging the basis. We did not adjust the basis size, the cutoff radius, or the weights for any of the results reported here.

\Cref{subfig:few-hmix} shows the mixing enthalpy of the disordered solid solution. Both A-atom potentials give a negative mixing enthalpy across the composition range, with a minimum near the equiatomic composition. The ACE-derived A-atom potential reaches $-79$\,meV/atom at 50\,at.\% W, against $-86$\,meV/atom for the EAM A-atom potential, and the two agree to within 9\,meV/atom at every composition. The Fe-W training dataset contained no large cells with random decorations and no configurations with appreciable chemical disorder. The mixing enthalpy of the disordered phase was therefore recovered entirely from the chemical interactions sampled in decorations of up to six atoms. The random supercell lies slightly closer to the ACE-derived A-atom potential than to the EAM A-atom potential, although these differences are small compared with the mixing enthalpy itself.

Having established that the bulk properties of the disordered phase are reproduced, we turn to line and planar defects. Peierls barriers in \cref{subfig:few-peierls} are reported only up to 50\,at.\% W, because at higher W content the EAM potential predicts an unphysical degenerate core structure for a screw dislocation placed at the easy core position. Up to 40\,at.\% W, both A-atom potentials predict barriers between 6.7 and 12.4\,meV/$b$, and both barriers increase towards the equiatomic composition. The difference between the two A-atom potentials is at most 5.9\,meV/$b$, which is small on the scale of the energy differences entering the fit, and we therefore do not interpret it as meaningful. The two A-atom potentials agree more closely for the $(110)$ surface energy (\cref{subfig:few-gamma110}). They give nearly identical values for pure Fe and W and differ by at most 0.17\,\unit{\joule\per\square\metre} in between, with the ACE-derived values higher at every intermediate composition. The rule-of-mixtures (ROM) estimate interpolates linearly between the surface energies of pure Fe and W. Both A-atom potentials show a positive deviation from this estimate.

\subsection{Mo-Nb from electronic structure calculations}
\label{sec:mo-nb-benchmark}
\begin{figure}
  \centering
  \subfloat[]{\label{subfig:monb-c11}\includegraphics[width=0.49\columnwidth]{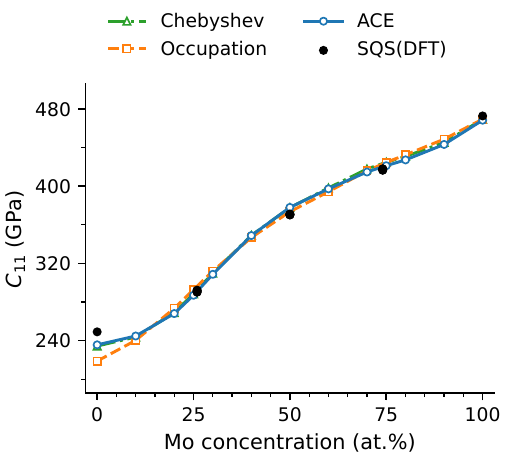}}\hfill
  \subfloat[]{\label{subfig:monb-c12}\includegraphics[width=0.49\columnwidth]{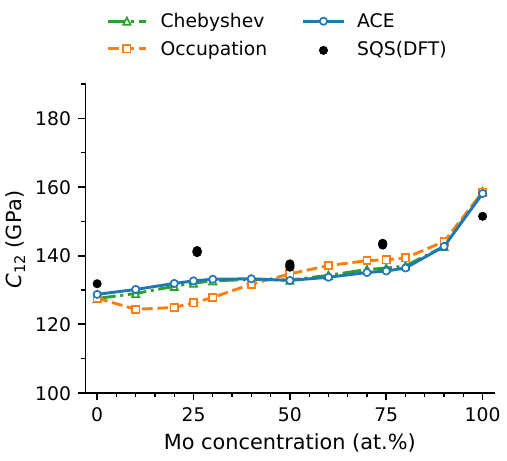}}\\
  \subfloat[]{\label{subfig:monb-c44}\includegraphics[width=0.49\columnwidth]{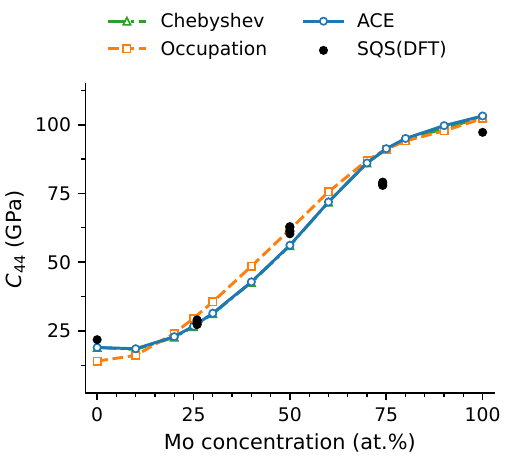}}\hfill
  \subfloat[]{\label{subfig:monb-lattice}\includegraphics[width=0.49\columnwidth]{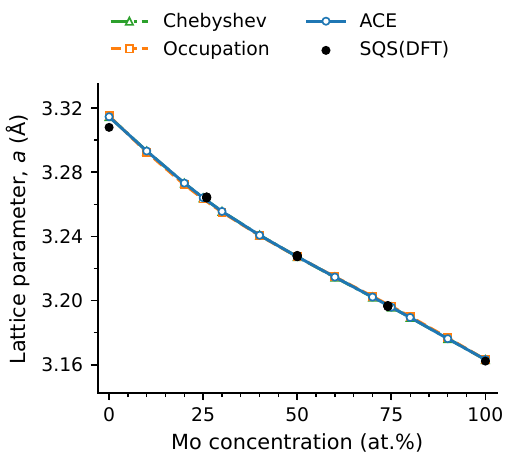}}\\
  \subfloat[]{\label{subfig:monb-hmix}\includegraphics[width=0.49\columnwidth]{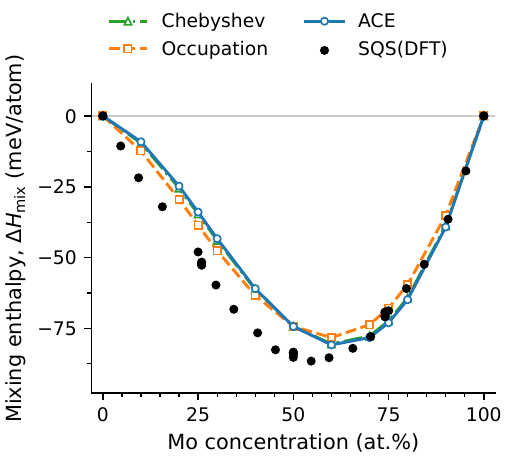}}
  \caption{Bulk properties of the disordered Mo-Nb solid solution as a function of composition, computed with A-atom potentials derived from linear ACE models using the Chebyshev, occupation, and ACE chemical site bases of \cref{eq:chemical_site_basis}. Filled symbols are DFT reference values, computed with special quasirandom structures for the alloy. (a) Elastic constant $C_{11}$. (b) Elastic constant $C_{12}$. (c) Elastic constant $C_{44}$. (d) Equilibrium lattice parameter of the bcc solid solution. (e) Mixing enthalpy.}
  \label{fig:mo-nb-bulk}
\end{figure}

We now turn to a more practical use of the A-atom potential, in which a linear ACE fitted to a small \textit{ab initio} training dataset is transformed into an A-atom potential to predict the properties of the disordered phase of a concentrated alloy. We do not aim for a precisely tuned ACE that captures detailed atomistic interactions such as short-range order or the sequence of ordered phases, and therefore restrict our benchmarks to properties of the disordered phase. We use the Mo-Nb binary alloy because the properties of the pure elements and of the disordered phase in this system have been computed from first principles in several studies\cite{weinberger_2013_peierlspotential,borges_2025_tunableenergy,liyanage_2026_crackingcase,ma_2019_effectstress,tran_2016_surfaceenergies,xu_2020_frankreadsource}. The linear ACE for Mo-Nb was parameterized on the training dataset of \cref{sec:training-dataset}, giving root-mean-square errors of $2$\,meV/atom in energy and $20$\,meV/\si{\angstrom} in force. We first benchmark the resulting A-atom potential against properties of the disordered phase that can be computed directly with DFT using SQS as approximants of the random alloy.

\Cref{fig:mo-nb-bulk} compares the elastic constants, lattice parameter, and mixing enthalpy computed with DFT for SQS against the predictions of the A-atom potential. The elastic constants in \crefrange{subfig:monb-c11}{subfig:monb-c44} are nearly identical for the Chebyshev, occupation, and ACE chemical site bases, and all three follow the composition dependence of the DFT values across the full range. The lattice parameter in \cref{subfig:monb-lattice} is reproduced with negligible differences, and the three bases are again indistinguishable. \Cref{subfig:monb-hmix} compares the mixing enthalpy of the disordered phase. Both the SQS and the A-atom potential give a negative mixing enthalpy that is asymmetric about the equiatomic composition, with its minimum on the Mo-rich side. The largest differences occur on the Nb-rich side, where the A-atom potential gives a less negative mixing enthalpy, and the SQS and A-atom values differ by at most about 15\,meV/atom. These differences are within the accuracy expected of the SQS approximation and of the fit of the linear ACE.

\begin{figure}
  \centering
  \subfloat[]{\label{subfig:monb-bain-nb}\includegraphics[width=0.49\columnwidth]{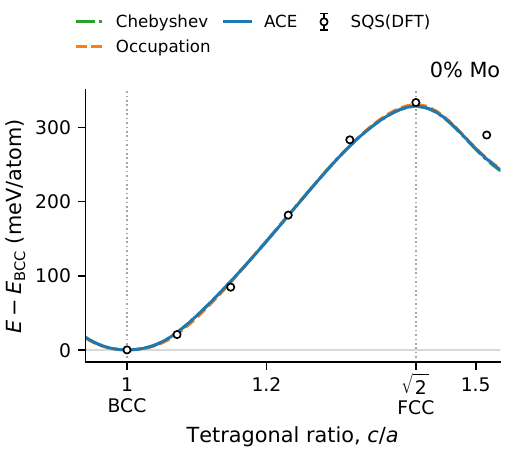}}\hfill
  \subfloat[]{\label{subfig:monb-bain-50}\includegraphics[width=0.49\columnwidth]{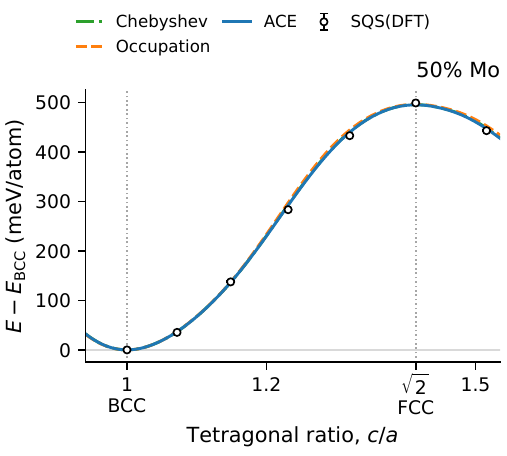}}\\
  \subfloat[]{\label{subfig:monb-bain-mo}\includegraphics[width=0.49\columnwidth]{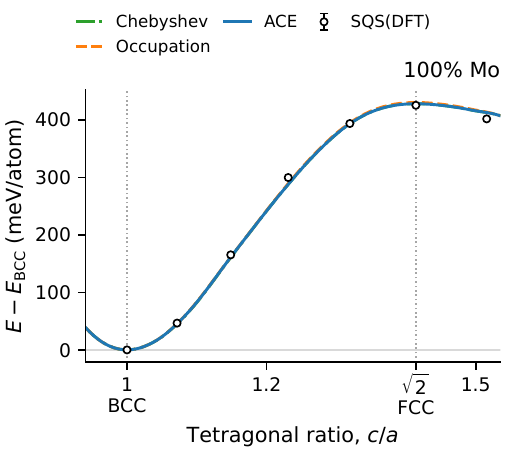}}
  \caption{Energy along the tetragonal Bain path of the disordered Mo-Nb solid solution, measured relative to the bcc structure and plotted against the tetragonal ratio $c/a$, where $c/a = 1$ is the bcc structure and $c/a = \sqrt{2}$ the fcc structure. Curves show A-atom potentials derived from linear ACE models using the Chebyshev, occupation, and ACE chemical site bases. Open symbols are DFT calculations, computed with special quasirandom structures for the alloy. (a) Pure Nb. (b) 50\,at.\% Mo. (c) Pure Mo.}
  \label{fig:mo-nb-bain}
\end{figure}

Given the agreement obtained for these bulk properties, we next compare the predictions of the A-atom potential for a structural phase transformation. The Bain path connects the bcc and fcc structures through a tetragonal distortion\cite{edgarbain_1924_naturemartensite,natarajan_2019_understandingdeformation,kolli_2020_discoveringhierarchies,kolli_2021_sixnew}. Along this path we compare the relative energies of the bcc and fcc structures, and check that bcc sits at a minimum and fcc at a maximum. No close-packed phase is expected to be stable in the Mo-Nb alloy, so an A-atom potential that stabilizes fcc would be of little use.

\Cref{fig:mo-nb-bain} compares the energies along the Bain path for pure Mo, pure Nb, and the equiatomic alloy. At all three compositions, both the A-atom potential and the DFT calculations place bcc at a minimum and fcc at a maximum, where it is mechanically unstable with respect to the tetragonal distortion. The energies along the path agree quantitatively with DFT for all three chemical site bases. The energy difference between fcc and bcc is larger for pure Mo than for pure Nb, and larger still for the equiatomic alloy. The A-atom potential reproduces this elevated value at the equiatomic composition, which is likely learned from the 57 fcc decorations included in the training dataset. The energy landscape of the disordered phase along the Bain path is thus reproduced from a small training dataset.

\begin{figure}
  \centering
  \subfloat[]{\label{subfig:monb-gamma100}\includegraphics[width=0.49\columnwidth]{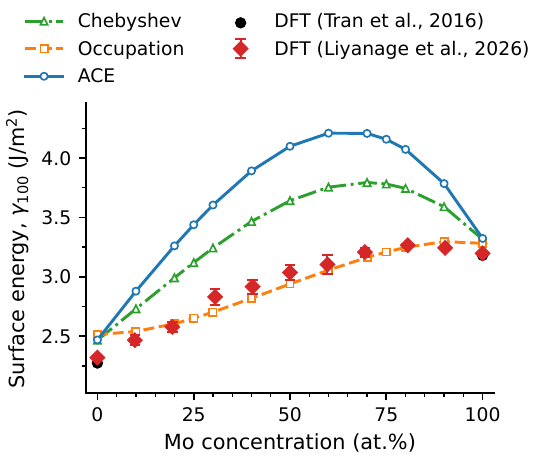}}\hfill
  \subfloat[]{\label{subfig:monb-gamma110}\includegraphics[width=0.49\columnwidth]{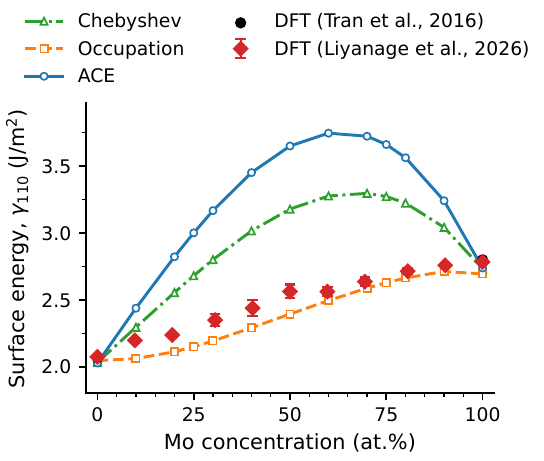}}\\
  \subfloat[]{\label{subfig:monb-usf110}\includegraphics[width=0.49\columnwidth]{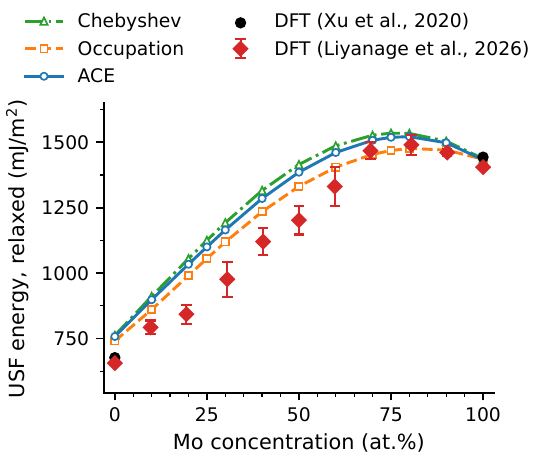}}\hfill
  \subfloat[]{\label{subfig:monb-usf112}\includegraphics[width=0.49\columnwidth]{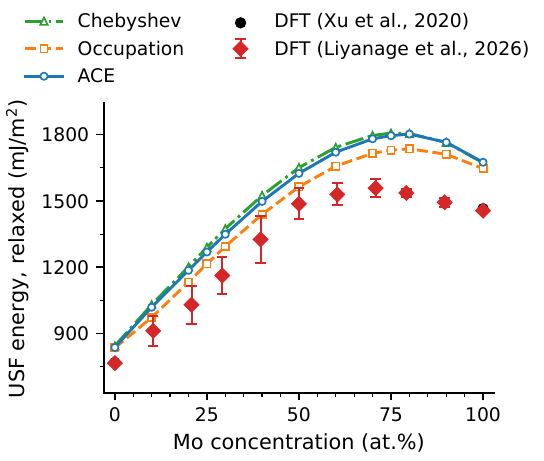}}
  \caption{Planar defect energies of the disordered Mo-Nb solid solution as a function of composition, computed with A-atom potentials derived from linear ACE models using the Chebyshev, occupation, and ACE chemical site bases. (a) Surface energy of the $(100)$ surface. (b) Surface energy of the $(110)$ surface. (c) Unstable stacking fault energy on $\{110\}$. (d) Unstable stacking fault energy on $\{112\}$. Reference values for the alloy are taken from Liyanage \textit{et al.}\cite{liyanage_2026_crackingcase}. Pure-element reference values are taken from Tran \textit{et al.}\cite{tran_2016_surfaceenergies} and Xu \textit{et al.}\cite{xu_2020_frankreadsource}.}
  \label{fig:mo-nb-planar}
\end{figure}

Beyond the comparisons against SQS, we now turn to properties for which the A-atom potential must extrapolate across the composition range between the two pure elements. \Cref{fig:mo-nb-planar} compares the predictions of the A-atom potential against DFT calculations from other studies for the surface and unstable stacking fault energies. The training dataset contained surfaces and unstable stacking faults only for pure Mo and Nb. \Cref{subfig:monb-gamma100,subfig:monb-gamma110} compare the surface energies against the calculations of Liyanage \textit{et al.}\cite{liyanage_2026_crackingcase}. The three chemical site bases give predictions that differ by up to 1.16\,\unit{\joule\per\square\metre} for $(100)$ and 1.26\,\unit{\joule\per\square\metre} for $(110)$ at intermediate compositions. The ACE and Chebyshev bases both predict large surface energies at intermediate compositions, while the occupation basis deviates only slightly from the ROM estimate. The spread in predictions from the three chemical site bases reflects extrapolation from the data-lean training set adopted here. The occupation basis closely follows the DFT values of Liyanage \textit{et al.}\cite{liyanage_2026_crackingcase} for both surfaces, which indicates that it may extrapolate material properties the most reliably of the three.

In contrast to the surface energies, the unstable stacking fault energies on the $\{110\}$ and $\{112\}$ planes vary much less between the three bases (\cref{subfig:monb-usf110,subfig:monb-usf112}). On $\{112\}$, the unstable stacking fault energy of pure Mo is over-predicted by approximately 13\%. All three bases lie above the DFT values of Liyanage \textit{et al.}\cite{liyanage_2026_crackingcase} at intermediate compositions, but reproduce the rise in the unstable stacking fault energy from pure Nb to a maximum at Mo-rich compositions.

\begin{figure}
  \centering
  \subfloat[]{\label{subfig:monb-vac-form}\includegraphics[width=0.8\columnwidth]{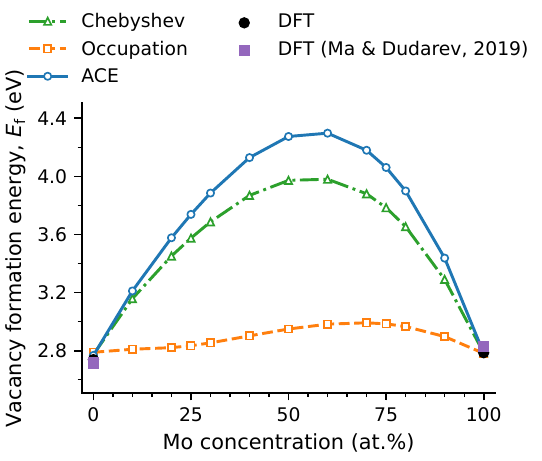}}\\
  \subfloat[]{\label{subfig:monb-vac-act}\includegraphics[width=0.8\columnwidth]{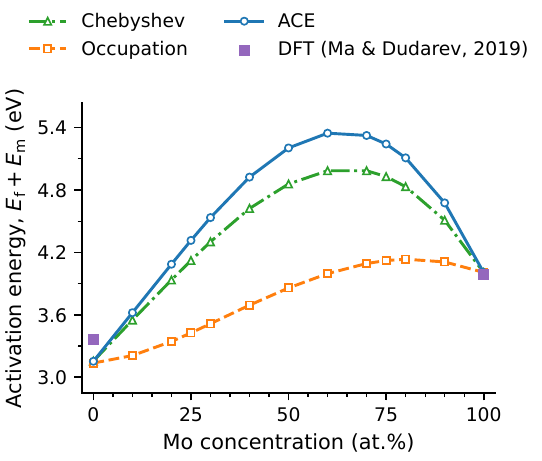}}
  \caption{Vacancy energetics of the disordered Mo-Nb solid solution as a function of composition, computed with A-atom potentials derived from linear ACE models using the Chebyshev, occupation, and ACE chemical site bases. (a) Vacancy formation energy $E_{\textrm{f}}$. (b) Activation energy for vacancy migration, $E_{\textrm{f}} + E_{\textrm{m}}$. Reference DFT values for the pure elements are taken from Ma and Dudarev\cite{ma_2019_effectstress}.}
  \label{fig:mo-nb-vacancy}
\end{figure}

\Cref{fig:mo-nb-vacancy} compares the vacancy formation and activation energies across the Mo-Nb composition range. As for the planar defects, the training dataset contained vacancies only in pure Mo and Nb. For the pure elements, the three chemical site bases agree with each other and with the DFT calculations of Ma and Dudarev\cite{ma_2019_effectstress}. The vacancy formation energy is similar for Mo and Nb (\cref{subfig:monb-vac-form}), while the activation energy is higher for pure Mo than for pure Nb (\cref{subfig:monb-vac-act}).

As with the surface energies, the three chemical site bases give different predictions for the vacancy properties at intermediate compositions (\cref{fig:mo-nb-vacancy}). The Chebyshev and ACE bases predict vacancy formation energies that rise to 1.2\,\unit{\electronvolt} and 1.5\,\unit{\electronvolt}, respectively, above the ROM estimate. In contrast, the occupation basis predicts a value at most 0.2\,\unit{\electronvolt} above the ROM estimate. The occupation basis is again the most likely of the three to represent the perfectly disordered phase. This is harder to test for vacancies, because DFT calculations of vacancy formation energies in concentrated alloys must account for the exchange chemical potentials of the elements\cite{lee_2026_modelingequilibrium}. Activation energies for vacancy migration show similar trends, rising 1.3\,\unit{\electronvolt} and 1.7\,\unit{\electronvolt} above the ROM estimate for the Chebyshev and ACE bases, respectively (\cref{subfig:monb-vac-act}). The occupation basis stays within 0.35\,\unit{\electronvolt} of the ROM estimate.

\begin{figure}
  \centering
  \subfloat[]{\label{subfig:monb-screw-core}\includegraphics[width=0.8\columnwidth]{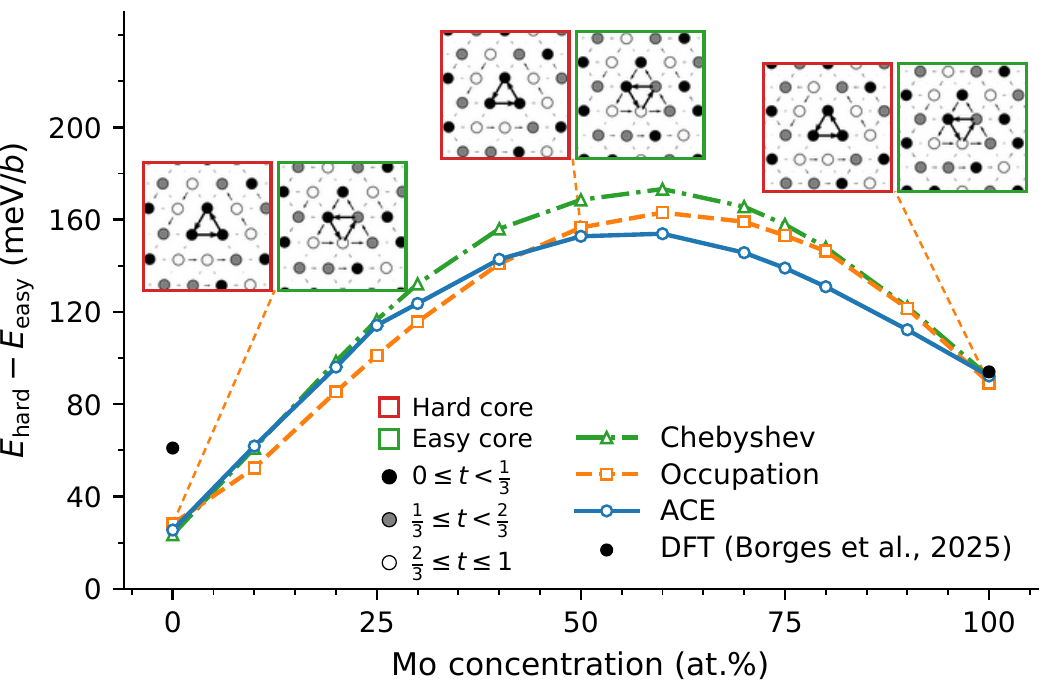}}\\
  \subfloat[]{\label{subfig:monb-screw-neb}\includegraphics[width=0.8\columnwidth]{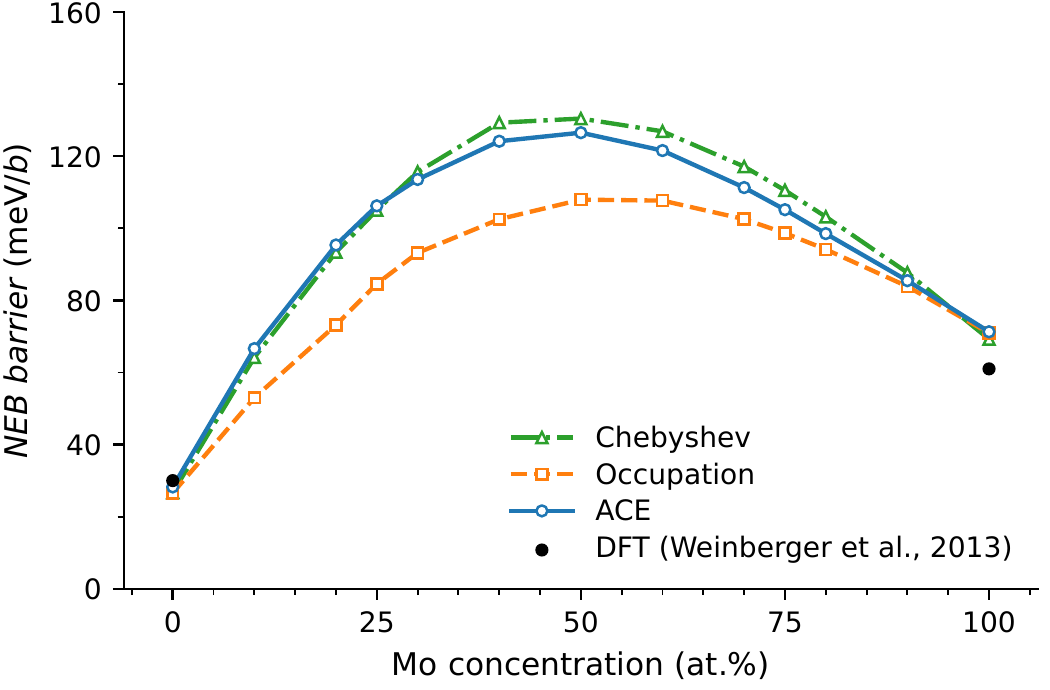}}
  \caption{Screw dislocation energetics of the disordered Mo-Nb solid solution as a function of composition, computed with A-atom potentials derived from linear ACE models using the Chebyshev, occupation, and ACE chemical site bases. (a) Energy difference between the hard and easy core configurations. Insets show differential displacement maps of the relaxed hard (red frame) and easy (green frame) cores obtained with the occupation-basis potential for Nb, Mo$_{50}$Nb$_{50}$ and Mo. Arrows represent the screw component of the relative displacement between neighboring $\langle 111 \rangle$ atomic columns, and atoms are shaded by their relaxed position along the dislocation line within one Burgers vector, $t \in [0,1)$, divided into thirds. (b) Peierls barrier obtained from nudged elastic band calculations, shown for a dipole. Reference electronic structure values for the pure elements are taken from Borges \textit{et al.}\cite{borges_2025_tunableenergy} and Weinberger \textit{et al.}\cite{weinberger_2013_peierlspotential}.}
  \label{fig:mo-nb-screw}
\end{figure}

Finally, we turn to the barriers for dislocation motion in the concentrated alloy. \Cref{fig:mo-nb-screw} compares two metrics that are commonly used to estimate the barrier to screw dislocation motion: the energy difference between core structures and the Peierls barrier. The energy difference between the hard and easy cores of a screw dislocation in bcc is the easier of the two to compute. \Cref{subfig:monb-screw-core} shows that the A-atom potential gives a positive value at every composition, placing the hard core above the easy core as expected from DFT calculations. For pure Mo, the prediction matches the DFT calculations of Borges \textit{et al.}\cite{borges_2025_tunableenergy}, while for pure Nb it is lower by $\approx$40\,meV/$b$. Higher training weights and additional training data did not reduce this difference. Borges \textit{et al.}\cite{borges_2025_tunableenergy} used a periodic screw-dipole configuration in cells of 135 atoms, while the A-atom predictions in \cref{subfig:monb-screw-core} use a dislocation monopole in cells of about 2000 atoms. The discrepancy for Nb may therefore arise from size effects, from the difference in cell geometry, or from errors in the linear ACE for pure Nb. At intermediate compositions, the energy difference between the hard and easy cores shows a positive deviation from the ROM estimate, and the three chemical site bases predict similar maxima of about 160\,meV/$b$. The A-atom potential also gives physically realistic core structures, as shown by the differential displacement maps in the insets of \cref{subfig:monb-screw-core}. The easy core has the expected non-degenerate structure across the full composition range.

The Peierls barrier in \cref{subfig:monb-screw-neb} is the maximum energy along the minimum-energy path computed with the nudged elastic band (NEB) method. It agrees more closely with DFT than the energy difference between the hard and easy cores. For both pure Mo and pure Nb, the A-atom potential reproduces the DFT calculations of Weinberger \textit{et al.}\cite{weinberger_2013_peierlspotential}. As with the energy difference between the hard and easy cores, the Peierls barrier shows a positive deviation from the ROM estimate, and reaches a maximum between 108 and 130\,meV/$b$ near the equiatomic composition. Since the lattice parameter varies only modestly across the Mo-Nb composition range (\cref{subfig:monb-lattice}), the Peierls stress of the perfectly disordered phase is also largest near the equiatomic composition.

\section{Discussion}
\label{sec:discussion}

The benchmarks of \cref{sec:fe-w-benchmark,sec:mo-nb-benchmark} show that an A-atom potential derived from a linear ACE describes the mixing enthalpy and the point, line, and planar defect energies of the disordered phase. Of the three chemical site bases, the occupation basis extrapolates most reliably to alloy compositions absent from the training dataset.

\begin{figure*}[!ht]
    \centering
    \includegraphics[width=0.9\linewidth]{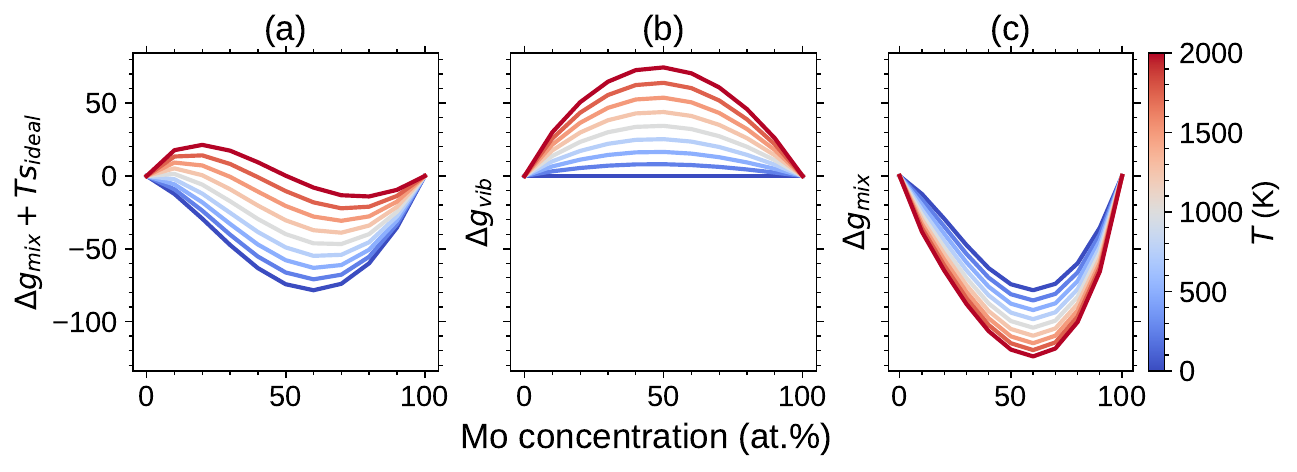}
    \caption{Quasi-harmonic free energies of the disordered Mo-Nb solid solution as a function of Mo concentration, computed with the A-atom potential derived from the occupation chemical site basis. Line colors indicate the temperature, from 0\,K to 2000\,K. All free energies are in meV/atom and are measured relative to pure Mo and Nb at the same temperature. (a) Mixing free energy excluding the ideal solution entropy, $\Delta g_{\mathrm{mix}} + T s_{\mathrm{ideal}}$. (b) Vibrational contribution to the mixing free energy, $\Delta g_{\mathrm{vib}}$, given by the difference between $\Delta g_{\mathrm{mix}} + T s_{\mathrm{ideal}}$ at temperature $T$ and its value at 0\,K. (c) Mixing free energy including the ideal solution entropy, $\Delta g_{\mathrm{mix}}$.}
    \label{fig:free_energy_MoNb}
\end{figure*}

The accuracy of a linear ACE depends on the size of its training dataset. To explore this dependence, we fitted linear ACE models to Mo-Nb training datasets of several sizes and compared the predicted properties for each. Smaller training datasets reproduced the material properties of the disordered phase, provided the chemical site basis was chosen carefully. As an example, we constructed a simpler model by retaining 750 of the 1273 strained structures, which reduces the Mo-Nb training dataset from 1722 to 1199 energies, and repeated the benchmarks of \cref{sec:mo-nb-benchmark} with it (\cref{supp-sec:training_reduction}). For the occupation basis, most properties are essentially unchanged. The main sensitivity is in the elastic response, where the mean error in $C_{12}$ increases from 7 to $11\,\unit{\giga\pascal}$ and that in $C_{44}$ from 5 to $10\,\unit{\giga\pascal}$, while $C_{11}$ is largely unchanged. The Chebyshev and ACE bases depend more strongly on the amount of training data, particularly for defect properties at intermediate compositions. With more training data, the three chemical site bases are expected to give similar predictions, as for the data-rich Fe-W dataset (\cref{fig:fe-w-benchmark}). The occupation-based A-atom potential thus retains its accuracy for most bulk and defect properties even after a substantial reduction in the training dataset.

Although the training dataset contained vacancies only in pure Mo and Nb, the occupation basis predicts vacancy formation energies above the ROM estimate at intermediate compositions (\cref{subfig:monb-vac-form}). Forming a vacancy in the disordered alloy is therefore more expensive than the vacancy formation energies of the pure elements would suggest. A similar positive deviation from the ROM estimate has been reported previously\cite{lee_2026_modelingequilibrium}. It can be related to the negative formation energies of ordered Mo-Nb phases and to the negative mixing enthalpy of the disordered phase (\cref{subfig:monb-hmix}), both of which indicate that Mo-Nb bonds are energetically favorable. Removing an atom breaks these bonds, which raises the vacancy formation energy above the ROM estimate. Since the A-atom potential describes a perfectly random alloy without short-range order, this positive deviation is intrinsic to the disordered phase rather than a consequence of local chemical ordering. The higher vacancy formation and activation energies (\cref{fig:mo-nb-vacancy}) would make vacancy-mediated diffusion in the alloy intrinsically ``sluggish'' relative to the ROM estimate.

The same positive deviation from the ROM estimate appears in the unstable stacking fault energies, the vacancy energies, and the barriers for screw dislocation motion, as well as in the fcc--bcc energy difference along the Bain path. None of these properties follows simple ROM behavior. The ROM estimate may not be the best baseline for judging whether an alloy has properties that differ from those expected of simple mixing. An A-atom potential provides a better baseline, because it includes the chemical interactions of the perfectly random alloy while excluding short-range order and local relaxations. Departures of a real alloy from the A-atom prediction can then be attributed to these effects.

In addition to static defect properties at 0\,K, the A-atom potential provides composition-dependent free energies of the perfectly disordered phase at finite temperature. Since the A-atom potential acts as the interatomic potential of an effective single-element material, these free energies can be computed without Monte Carlo sampling over chemical configurations. The free energy of the disordered phase can then be compared with those of ordered phases to estimate the order--disorder transition temperature. \Cref{fig:free_energy_MoNb} shows the quasi-harmonic mixing free energy of Mo-Nb as a function of Mo concentration at temperatures from 0\,K to 2000\,K. At each temperature, the mixing free energy is measured relative to the free energies of pure Mo and Nb at that same temperature. The free energies of the alloy and of the pure elements all decrease with temperature relative to a common reference state, while the mixing free energy can increase. To focus on the enthalpic and vibrational contributions, the free energy in \cref{fig:free_energy_MoNb}(a) excludes the ideal solution entropy. At 0\,K this free energy equals the mixing enthalpy of \cref{subfig:monb-hmix}, and it increases with temperature, becoming positive on the Nb-rich side at the highest temperatures. The vibrational free energy of the alloy therefore decreases more slowly with temperature than the composition-weighted average of those of the pure elements.

Subtracting the 0\,K mixing enthalpy from the free energies in \cref{fig:free_energy_MoNb}(a) gives the vibrational contribution to the mixing free energy, shown in \cref{fig:free_energy_MoNb}(b). This contribution is positive and increases with temperature, so lattice vibrations destabilize the disordered phase relative to the pure elements. Phase stability, however, depends on the balance between configurational and vibrational contributions. Including the ideal solution entropy more than compensates for the vibrational contribution (\cref{fig:free_energy_MoNb}(c)), and the mixing free energy of the disordered phase becomes more negative with increasing temperature. For phase diagrams, the relevant vibrational quantity is the difference in vibrational free energy between the ordered and disordered phases, and its size depends on the alloy. For Ni$_3$Al, van de Walle \textit{et al.}\cite{vandewalle_1998_firstprinciplescomputation} found this difference to be nearly zero within the harmonic approximation. A careful comparison of the vibrational free energies of ordered and disordered phases is left for future work. The A-atom potential developed here provides a direct route to the finite-temperature free energy of the perfectly disordered phase for such comparisons.

\section{Conclusion}
\label{sec:conclusion}

We have benchmarked A-atom potentials derived exactly from linear ACE models. For Fe-W, the A-atom potential derived from a linear ACE agrees with the A-atom potential obtained directly from the parent EAM potential. For Mo-Nb, about 1500 electronic structure calculations of energies and forces were sufficient to parameterize a linear ACE whose A-atom potential agrees with DFT calculations for SQS and with DFT data from the literature. This A-atom potential predicts physically realistic screw dislocation cores and the variation of defect properties with alloy composition. Of the three chemical site bases, the occupation basis gave the most reliable predictions at alloy compositions absent from the training dataset. The A-atom potential also gives direct access to the finite-temperature free energy of the disordered phase. For Mo-Nb, the vibrational contribution to the mixing free energy destabilizes the perfectly disordered phase, and this destabilization grows with temperature. The ideal solution entropy outweighs the vibrational contribution, and the disordered phase is stabilized overall. A-atom potentials derived from linear ACE models can therefore serve as a baseline for quantifying the effects of short-range order and other departures from ideal randomness, and for estimating the finite-temperature properties of disordered phases.

\section*{Data availability}
The data that support the findings of this study are openly available on Materials Cloud at [LINK].

\begin{acknowledgments}
  This research was supported by the Swiss National Science Foundation (grant number 215178). ARN is grateful for support through the framework of the EUROfusion Consortium, partially funded by the European Union via the Euratom Research and Training Programme (grant agreement no. 101052200---EUROfusion). Views and opinions expressed are however those of the author(s) only and do not necessarily reflect those of the European Union or the European Commission. Neither the European Union nor the European Commission can be held responsible for them.
\end{acknowledgments}
\appendix

\section{Composition-averaged ZBL repulsion}
\label{app:average_zbl}

The Ziegler--Biersack--Littmark (ZBL) potential describes the screened nuclear repulsion between two atoms and is used here as an explicit short-range repulsive baseline. For a pair of chemical species $p$ and $q$, the interaction is
\begin{equation}
  E_{pq}^{\mathrm{ZBL}}(r)
  =
  \frac{e^2}{4\pi\varepsilon_0}
  \frac{Z_p Z_q}{r}
  \phi\left(\frac{r}{a_{pq}}\right),
  \label{eq:zbl}
\end{equation}
where $Z_p$ and $Z_q$ are the atomic numbers of the two species, $\phi$ is the universal ZBL screening function, and $a_{pq}$ is the screening length for the pair. During training of the linear ACE, the species-resolved ZBL energies and forces are subtracted from the reference data, and the linear ACE is fitted to the remaining contribution.

In the linear ACE, the species on each site is known, and the appropriate $AA$, $AB$, or $BB$ ZBL interaction can be evaluated directly. This is no longer possible once the linear ACE is transformed into an A-atom potential with any of the chemical site bases, because each site is then occupied by an average atom that represents a composition rather than a definite species. Assigning a single atomic number to the average atom would not provide the appropriate short-range interaction.

Since the ZBL energy is a pair interaction, its A-atom average is obtained in the same way as for any pair potential. For a binary alloy $A_{1-x}B_x$ with random occupations, the ZBL contribution is averaged over the possible pair occupations:
\begin{equation}
  \begin{split}
    \overline{E}_{\mathrm{ZBL}}(r;x)
    ={} &
    (1-x)^2 E_{AA}^{\mathrm{ZBL}}(r)
    \\
        & + 2x(1-x) E_{AB}^{\mathrm{ZBL}}(r)
    \\
        & + x^2 E_{BB}^{\mathrm{ZBL}}(r),
  \end{split}
  \label{eq:average_zbl}
\end{equation}
where $x$ is the concentration of species $B$. The weights are the probabilities of $AA$, $AB$ or $BA$, and $BB$ pairs in a random binary alloy. The composition-averaged interaction in \cref{eq:average_zbl} replaces the species-resolved ZBL term and is added to the A-atom potential as its short-range repulsive baseline.

\bibliographystyle{apsrev4-2}
\bibliography{bibliography}

\end{document}


\title{Supplementary Information for ``Benchmarking average atom potentials derived from atomic cluster expansions''}

\author{Deepak Somani}
\affiliation{Laboratory of Materials Design and Simulation (MADES), Institute of Materials, \'{E}cole Polytechnique F\'{e}d\'{e}rale de Lausanne}

\author{Lorenzo Piersante}
\affiliation{Laboratory of Materials Design and Simulation (MADES), Institute of Materials, \'{E}cole Polytechnique F\'{e}d\'{e}rale de Lausanne}

\author{Anirudh Raju Natarajan}
\email{anirudh.natarajan@epfl.ch}
\affiliation{Laboratory of Materials Design and Simulation (MADES), Institute of Materials, \'{E}cole Polytechnique F\'{e}d\'{e}rale de Lausanne}

\maketitle

\section{Mo-Nb benchmarks with a smaller training dataset}
\label{sec:training_reduction}

\begin{figure}
    \centering

    \subfloat[]{%
        \label{subfig:monb_reduced-c11}%
        \includegraphics[width=0.32\linewidth]{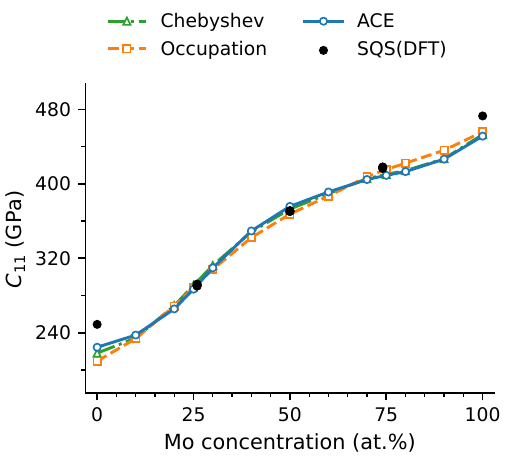}%
    }\hfill
    \subfloat[]{%
        \label{subfig:monb_reduced-c12}%
        \includegraphics[width=0.32\linewidth]{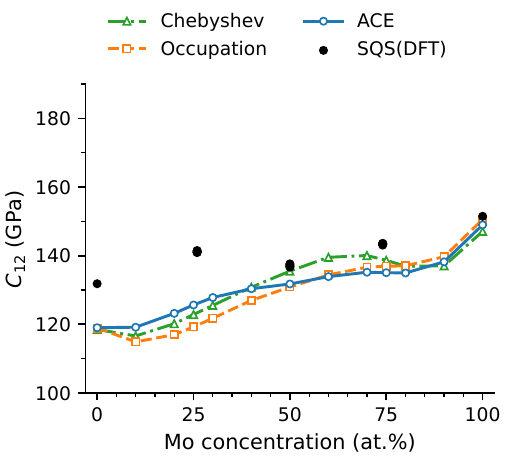}%
    }\hfill
    \subfloat[]{%
        \label{subfig:monb_reduced-c44}%
        \includegraphics[width=0.32\linewidth]{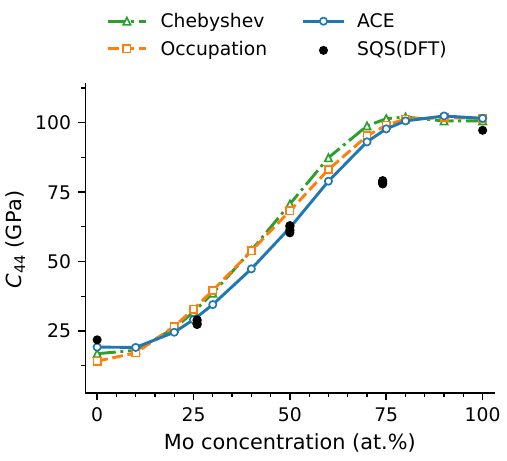}%
    }\\[2mm]

    \makebox[\linewidth][c]{%
        \subfloat[]{%
            \label{subfig:monb_reduced-lattice}%
            \includegraphics[width=0.32\linewidth]{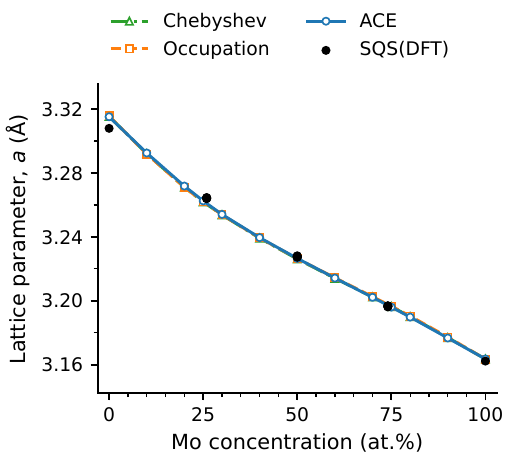}%
        }\hspace{0.02\linewidth}
        \subfloat[]{%
            \label{subfig:monb_reduced-hmix}%
            \includegraphics[width=0.32\linewidth]{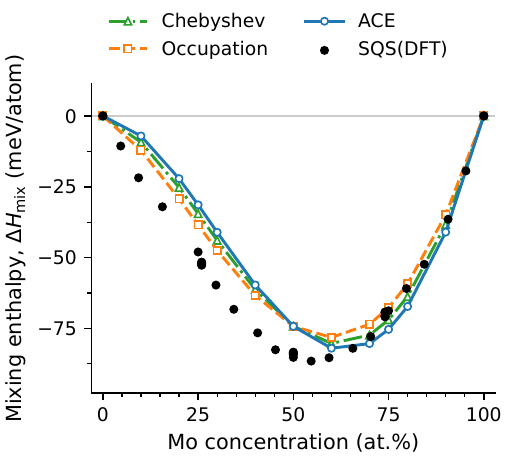}%
        }%
    }

    \caption{Bulk properties of the disordered Mo-Nb solid solution as a function of composition, computed with A-atom potentials derived from linear ACE models fitted to the smaller training dataset of 941 calculations, using the Chebyshev, occupation, and ACE chemical site bases. Filled symbols are DFT reference values, computed with special quasirandom structures for the alloy. (a) Elastic constant $C_{11}$. (b) Elastic constant $C_{12}$. (c) Elastic constant $C_{44}$. (d) Equilibrium lattice parameter of the bcc solid solution. (e) Mixing enthalpy.}
    \label{fig:mo-nb-reduced-bulk}
\end{figure}

\begin{figure}
    \centering

    \subfloat[]{%
        \label{subfig:monb_reduced-bain-nb}%
        \includegraphics[width=0.32\linewidth]{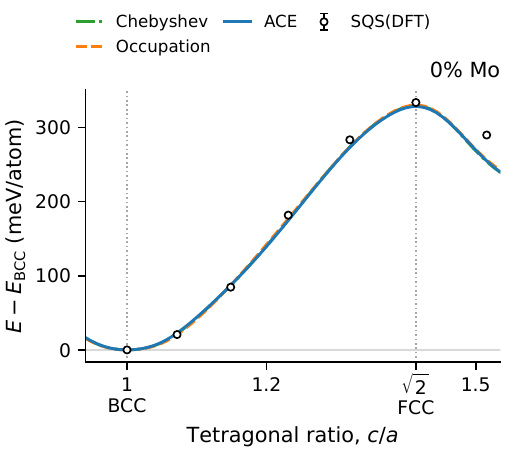}%
    }\hfill
    \subfloat[]{%
        \label{subfig:monb_reduced-bain-50}%
        \includegraphics[width=0.32\linewidth]{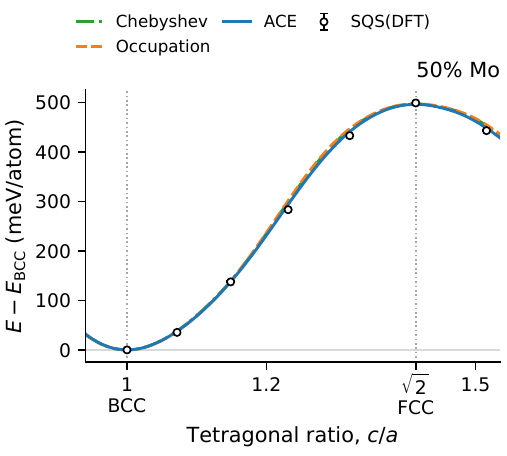}%
    }\hfill
    \subfloat[]{%
        \label{subfig:monb_reduced-bain-mo}%
        \includegraphics[width=0.32\linewidth]{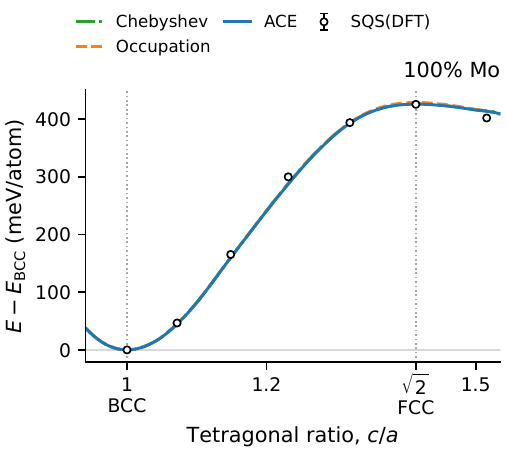}%
    }

    \caption{Energy along the tetragonal Bain path of the disordered Mo-Nb solid solution, measured relative to the bcc structure and plotted against the tetragonal ratio $c/a$, where $c/a = 1$ is the bcc structure and $c/a = \sqrt{2}$ the fcc structure. Curves show A-atom potentials derived from linear ACE models fitted to the smaller training dataset of 941 calculations, using the Chebyshev, occupation, and ACE chemical site bases. Open symbols are DFT calculations, computed with special quasirandom structures for the alloy. (a) Pure Nb. (b) 50\,at.\% Mo. (c) Pure Mo.}
    \label{fig:mo-nb-reduced-bain}
\end{figure}

\begin{figure}
    \centering

    \makebox[\linewidth][c]{%
        \subfloat[]{%
            \label{subfig:monb_reduced-gamma100}%
            \includegraphics[width=0.32\linewidth]{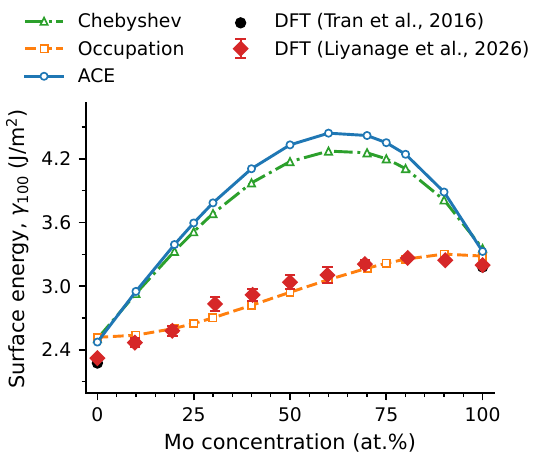}%
        }\hspace{0.03\linewidth}
        \subfloat[]{%
            \label{subfig:monb_reduced-gamma110}%
            \includegraphics[width=0.32\linewidth]{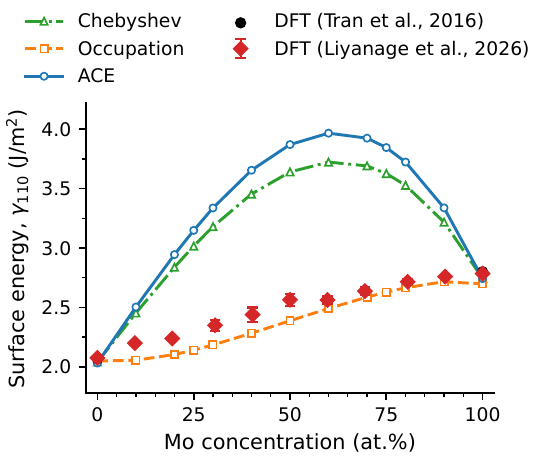}%
        }%
    }\\[2mm]

    \makebox[\linewidth][c]{%
        \subfloat[]{%
            \label{subfig:monb_reduced-usf110}%
            \includegraphics[width=0.32\linewidth]{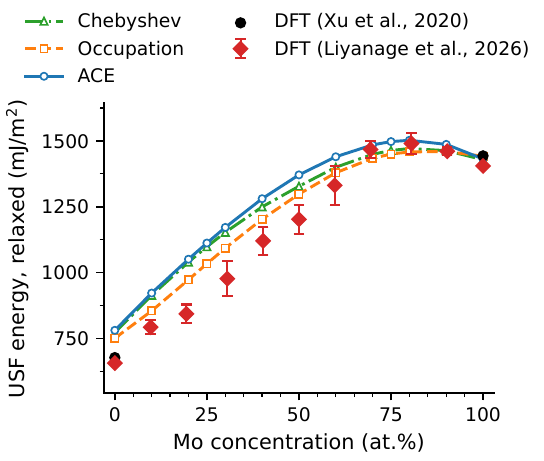}%
        }\hspace{0.03\linewidth}
        \subfloat[]{%
            \label{subfig:monb_reduced-usf112}%
            \includegraphics[width=0.32\linewidth]{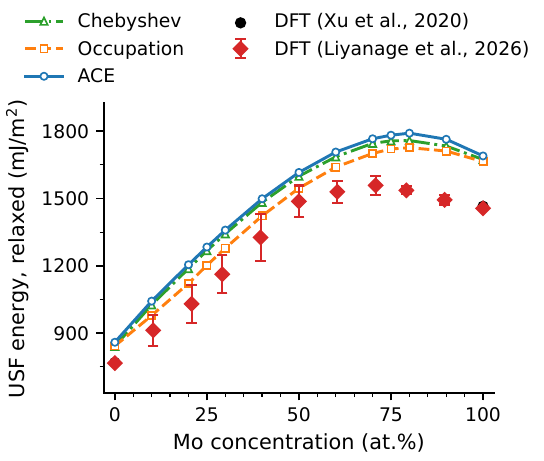}%
        }%
    }

    \caption{Planar defect energies of the disordered Mo-Nb solid solution as a function of composition, computed with A-atom potentials derived from linear ACE models fitted to the smaller training dataset of 941 calculations, using the Chebyshev, occupation, and ACE chemical site bases. (a) Surface energy of the $(100)$ surface. (b) Surface energy of the $(110)$ surface. (c) Unstable stacking fault energy on $\{110\}$. (d) Unstable stacking fault energy on $\{112\}$. Reference values for the alloy are taken from Liyanage \textit{et al.}\cite{liyanage_2026_crackingcase}. Pure-element reference values are taken from Tran \textit{et al.}\cite{tran_2016_surfaceenergies} and Xu \textit{et al.}\cite{xu_2020_frankreadsource}.}
    \label{fig:mo-nb-reduced-planar}
\end{figure}

\begin{figure}
    \centering

    \subfloat[]{%
        \label{subfig:monb_reduced-vac-form}%
        \includegraphics[width=0.32\linewidth]{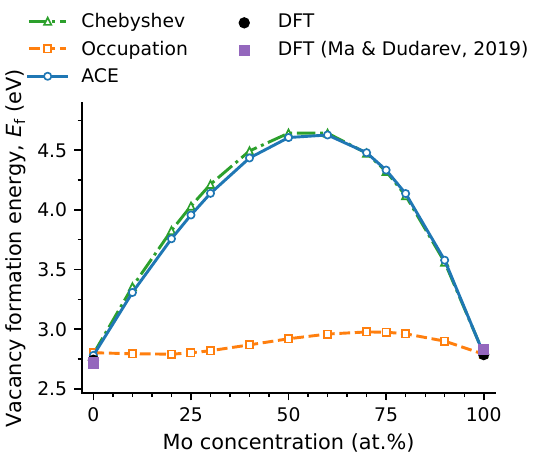}%
    }\hspace{0.03\linewidth}
    \subfloat[]{%
        \label{subfig:monb_reduced-vac-act}%
        \includegraphics[width=0.32\linewidth]{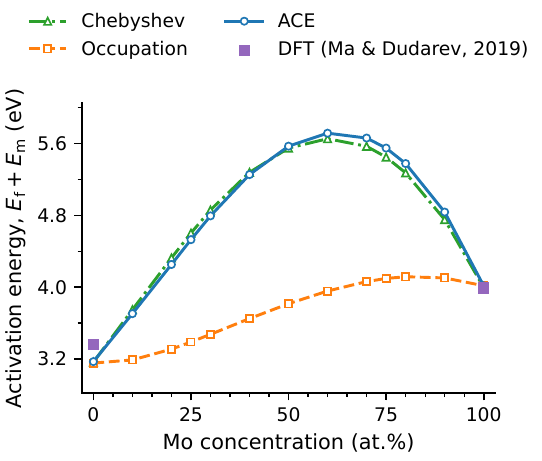}%
    }

    \caption{Vacancy energetics of the disordered Mo-Nb solid solution as a function of composition, computed with A-atom potentials derived from linear ACE models fitted to the smaller training dataset of 941 calculations, using the Chebyshev, occupation, and ACE chemical site bases. (a) Vacancy formation energy $E_{\textrm{f}}$. (b) Activation energy for vacancy migration, $E_{\textrm{f}} + E_{\textrm{m}}$. Reference DFT values for the pure elements are taken from Ma and Dudarev\cite{ma_2019_effectstress}.}
    \label{fig:mo-nb-reduced-vacancy}
\end{figure}

\begin{figure}
    \centering

    \subfloat[]{%
        \label{subfig:monb_reduced-screw-core}%
        \includegraphics[width=0.32\linewidth]{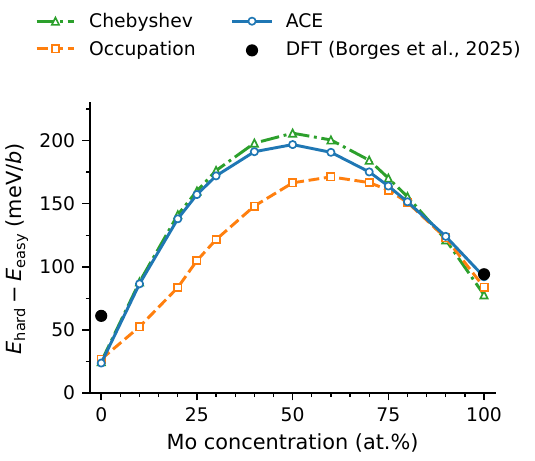}%
    }\hspace{0.03\linewidth}
    \subfloat[]{%
        \label{subfig:monb_reduced-screw-neb}%
        \includegraphics[width=0.32\linewidth]{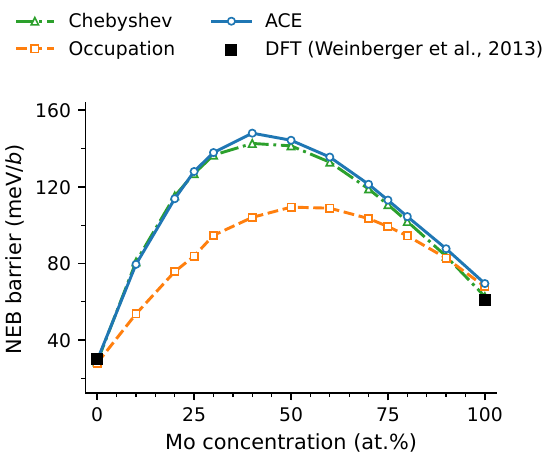}%
    }

    \caption{Screw dislocation energetics of the disordered Mo-Nb solid solution as a function of composition, computed with A-atom potentials derived from linear ACE models fitted to the smaller training dataset of 941 calculations, using the Chebyshev, occupation, and ACE chemical site bases. (a) Energy difference between the hard and easy core configurations. (b) Peierls barrier obtained from nudged elastic band calculations, shown for a dipole. Reference electronic structure values for the pure elements are taken from Borges \textit{et al.}\cite{borges_2025_tunableenergy} and Weinberger \textit{et al.}\cite{weinberger_2013_peierlspotential}.}
    \label{fig:mo-nb-reduced-screw}
\end{figure}

To gauge how strongly the predictions depend on the training data, we compared A-atom potentials derived from linear ACE models fitted to two Mo-Nb training datasets. The two datasets differ only in the number of strained structures retained from the reference dataset by the Mahalanobis-distance selection. The larger dataset used in the main text contains 1464 calculations, and the smaller dataset retains 750 of the 1273 strained structures, giving 941 calculations. \Cref{fig:mo-nb-reduced-bulk,fig:mo-nb-reduced-bain,fig:mo-nb-reduced-planar,fig:mo-nb-reduced-vacancy,fig:mo-nb-reduced-screw} show the bulk, Bain path, planar defect, vacancy, and screw dislocation properties predicted by the potentials fitted to the smaller dataset. The DFT and literature reference data are the same as in the main text. Only the training dataset used to construct the potentials is changed.

The smaller dataset gives predictions similar to those of the larger dataset. For the occupation basis, the predictions change little between the two datasets. The mixing enthalpy, for instance, changes by at most 0.24\,meV/atom, and the changes in the defect properties are similarly small. The elastic constants are the most sensitive of these properties, as discussed in the main text. The Chebyshev and ACE bases are more sensitive to the size of the training dataset, and their predicted properties can change by nearly 20\% relative to the larger dataset. This larger sensitivity at intermediate compositions indicates that the Chebyshev and ACE predictions are more strongly affected by reduced sampling of the combined composition and environment space. The occupation basis also shows the smallest deviations from the ROM estimate and remains comparatively insensitive to the reduction in training data.

\bibliographystyle{apsrev4-2}
\bibliography{bibliography}